\documentclass[aps,pre,twocolumn,showpacs,showkeys,superscriptaddress]{revtex4}

\usepackage{amsmath}
\usepackage{graphicx}
\usepackage{amssymb}

\usepackage{ulem}
\newcommand{\corr}[2]{\sout{#1}\uline{#2}}
 \renewcommand{\corr}[2]{#2}

\begin{document}
\title{Modulating coherence resonance in non-excitable systems by time-delayed feedback}

\author{Paul M. Geffert}
\affiliation{Institut f{\"u}r Theoretische Physik, Technische Universit{\"a}t Berlin,
Hardenbergstr.\ 36, D-10623 Berlin, Germany}
\author{Anna Zakharova}
\affiliation{Institut f{\"u}r Theoretische Physik, Technische Universit{\"a}t Berlin,
Hardenbergstr.\ 36, D-10623 Berlin, Germany}
\author{Andrea V{\"u}llings}
\affiliation{Institut f{\"u}r Theoretische Physik, Technische Universit{\"a}t Berlin,
Hardenbergstr.\ 36, D-10623 Berlin, Germany}
\author{Wolfram Just}
\email{w.just@qmul.ac.uk}
\affiliation{Queen Mary, University of London, School of Mathematical
Sciences, Mile End Road, London E1 4NS, UK}
\affiliation{Institut f{\"u}r Theoretische Physik, Technische Universit{\"a}t Berlin,
Hardenbergstr.\ 36, D-10623 Berlin, Germany}
\author{Eckehard Sch{\"o}ll}
\email{schoell@physik.tu-berlin.de}
\affiliation{Institut f{\"u}r Theoretische Physik, Technische Universit{\"a}t Berlin,
Hardenbergstr.\ 36, D-10623 Berlin, Germany}

\begin{abstract}
We propose a paradigmatic model system, a subcritical Hopf normal form subjected to noise and
time-delayed feedback, to investigate the impact of time delay on coherence resonance in
non-excitable systems. We develop analytical tools to estimate the stationary distribution and
the time correlations in nonlinear stochastic delay differential equations. These tools are applied
to our model to propose a novel quantity to \corr{explain the mechanism of}{measure} coherence resonance induced by a 
saddle-node bifurcation of periodic orbits. 
\end{abstract}

\pacs{ }
\keywords{ }

\maketitle

\section{Introduction}

The advent of nonlinear dynamics, originally a discipline in pure mathematics, in the
physical sciences at about three decades ago triggered an avalanche of new developments,
which still to date has not elapsed at all. In particular, notions like chaos and sensitive dependence
on initial conditions put new emphasis on the study how small perturbations change the dynamics 
of physical systems. That gave new stimulus to topics like the classical Kramers problem
of thermal activation \cite{KRA40}, which originated in physical chemistry, and which nowadays is considered
as one of the early milestones in the study of complex behaviour subjected to noise \cite{HAE90}.
In the wake of such renewed interest various quite counterintuitive discoveries have been made,
most notably the ability of weak noise to increase the regularity of a signal. Depending on the
underlying mechanism different terms have been coined for these phenomena. Stochastic resonance
refers to an increase of a signal-to-noise ratio which follows the spirit of the
original Kramers problem subjected to periodic driving \cite{GAM98}.
In the setting of a transport problem the concept of stochastic resonance
gave rise to new mechanisms for nonequilibrium transport \cite{REI02a}. 
The term coherence resonance refers to the nonmonotonic dependence of the coherence or regularity upon
noise intensity in an autonomous excitable system without external periodic driving, which 
leads to an optimum coherence of noise-induced oscillations at non-zero noise intensity
\cite{GAN93,PIK97,LIN04}.
Originally the term coherence resonance has been restricted to excitable systems, where the mechanism is based
upon the existence of two competing time-scales with opposite dependence upon noise intensity, producing a 
nonmonotonic overall dependence:
The excitable system rests in a locally stable steady state (rest state), and emits a spike upon excitation beyond a threshold, 
i.e., the dynamics performs a long excursion in phase space, before returning  to the rest state. With increasing noise the excitation across threshold occurs more frequently, and thus the interspike intervals become more regular. On the other hand, with increasing 
noise the deterministic spiking dynamics becomes smeared out, which counteracts the former effect. 
Such behaviour has been shown theoretically or experimentally in a
variety of excitable systems, like lasers with saturable absorber \cite{DUB99}, optical feedback \cite{GIA00,AVI04,OTT14a}, 
and optical injection \cite{ZIE13}, semiconductor superlattices \cite{HIZ06}, or neural systems \cite{PIK97,LIN04}. 
The mechanism of coherence resonance close to bifurcations 
has been investigated numerically, where characteristic signatures of noisy precursors of a dynamical instability appear 
in the power spectrum \cite{WIE85,NEI97}.

Here we want to focus on coherence resonance in non-excitable systems where the underlying
mechanism is associated with a subcritical Hopf bifurcation and a saddle-node bifurcation of a pair of (stable and unstable)
periodic orbits \cite{USH05,ZAK10a}.
The fundamental aspects and the mechanism of such a novel coherence resonance scenario
are elaborated in \cite{ZAK13}. \corr{}{Unlike coherence resonance in excitable systems, which is linked two competing timescales, 
the main mechanism of this novel resonance is related to a transient phase space structure - a ghost of a saddle-node bifurcation of periodic orbits. 
In this respect the coherence resonance mechanism has something in common with type I intermittent dynamics. 
But contrary to the latter, the deterministic dynamics in our case does not provide a reinjection mechanism which here solely depends on the stochastic forcing and thus gives rise to the resonance characteristics.} The essentially new element we want to add here
is the investigation of impact of time-delayed feedback. Dynamics with time delay, even though
well studied in engineering science \cite{BEL63}, has become only recently a rapidly developing
subject in applied nonlinear dynamical systems theory (see e.g. \cite{JUS09, FLU13} for recent accounts).
Previous studies of coherence resonance in the presence of time delay were restricted to excitable systems
where mainly numerical but also some analytical results were obtained \cite{JAN03,BAL04,PRA07,HIZ08b,AUS09,KOU10a}.
Delay-coupled Hopf normal forms have been studied in \cite{VUE14} with numerical and analytical methods using linear response theory.
Further progress in analytical terms is hampered by the observation that stochastic delay differential equations  
do not define a priori a Markov process, so that all the powerful tools like Fokker-Planck
equations are not available. There have been attempts to
reduce stochastic delay differential systems to Fokker-Planck equations, but these approaches are either
limited to linear systems, or they involve some quite severe approximations which are
difficult to control \cite{FRA03a}. Hence, we also consider the problem we are dealing with, 
the impact of time delay on coherence resonance, as opportunity to develop and to present
some of the analytical techniques that can be applied for the analysis of general
nonlinear stochastic time delay systems. While these techniques are 
variants of well established approaches in dynamical systems theory, their
application to nonlinear stochastic time delay dynamics is certainly
not common knowledge, and will be presented here in a coherent form.

\begin{figure}[h!]

\includegraphics[width=0.4 \textwidth]{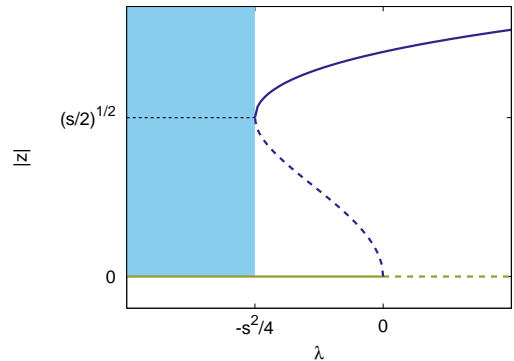}

\caption{Diagrammatic view of the bifurcation diagram of the deterministic
system without time delay, eq.(\ref{aa}) for $K=0$ and $D=0$, in the subcritical
regime $s>0$. Bronze (grey): trivial
fixed point (solid: stable, dashed: unstable), blue (black): periodic orbit 
(solid: stable, dashed: unstable), dotted line: location of the saddle-node bifurcation of 
periodic orbits. The shaded region is the parameter 
region of interest for coherence resonance.}
\label{figk}
\end{figure}

We focus on a paradigmatic model, a complex Hopf normal form
subjected to noise and time-delayed feedback
\begin{eqnarray}\label{aa}
\dot{z}(t)&=&(\lambda+i\omega +s|z(t)|^2-|z(t)|^4)z(t) \nonumber \\
& &-K(z(t)-z(t-\tau))+\sqrt{2D}\zeta(t) \, ,
\end{eqnarray}
where $z \in \mathbb{C}$, $\lambda \in \mathbb{R}$ and $s \in \mathbb{R}$ are the deterministic bifurcation parameters, $\omega$ denotes the intrinsic frequency of the system. 
$D \geq 0$ is the noise intensity, $K \in \mathbb{R}$ is the strength of the feedback, and $\tau$ is the delay time.
$\zeta(t)$ denotes normalised complex valued isotropic Gaussian white noise
$\langle \zeta(t) \zeta ^*(t')\rangle = 2 \delta(t-t')$, and we focus solely
on the subcritical case $s>0$. Then the system without time delay and without
noise, $K=0$ and $D=0$, shows the classic scenario of a subcritical Hopf bifurcation
with the creation of an unstable periodic orbit at $\lambda=0$,
bistability between the stable trivial fixed point and a stable periodic orbit in the 
range $-s^2/4<\lambda<0$, and the collision of stable and unstable periodic orbits at
$\lambda=-s^2/4$, see figure \ref{figk}. The parameter region $\lambda<-s^2/4$ 
is the region of interest from the point of view of coherence resonance. 
The time-delayed feedback implemented in 
the model, eq.(\ref{aa}), is similar to time-delayed feedback control of deterministic orbits 
\cite{PYR92}, which is known to be able to enhance and stabilise oscillatory 
behaviour (see e.g. \cite{SCH07} for a comprehensive account on this 
and related subjects). Hence, a feedback of such a type is an obvious choice 
to investigate the impact of time delay on coherence resonance.

We are now going to address two issues separately. In section \ref{sec2}
we will discuss in detail the properties of the stationary probability distribution
of eq.(\ref{aa}), in particular with regards to its dependence on the time delay.
We will develop an approach along the lines of a  centre manifold reduction, 
which allows us to derive closed analytical expressions for the
stationary probability distribution. Section \ref{sec3}
is devoted to the investigation of correlation properties of eq.(\ref{aa}),
shifting the focus to the core subject
of resonance properties of the underlying dynamics. Here as well we will
put the emphasis on \corr{an analytic}{a} scheme, which will allow us to compute 
analytically the
correlation function and the corresponding power spectrum with
high accuracy. Moreover, we will introduce a novel quantity - ghost weight - to \corr{explain}{quantify} coherence resonance that occurs, though not in a pronounced form, even outside the regime of bimodal stationary probability distributions. 
Finally, in section \ref{sec4} we \corr{will merge both aspects}{conclude with remarks how the ideas developed in the previous sections can be merged} for the benefit to
understand and characterise coherence resonance in stochastic time delay 
systems.
\corr{}{Some of the analytical techniques - the bifurcation analysis of the deterministic model and the centre manifold reduction of the stochastic equations - are summarised in two appendices, for the convenience of the reader.}

\section{Stationary behaviour and stochastic bifurcations} \label{sec2}

It is not common practise to look at
resonance phenomena in stochastic systems
from a traditional dynamical systems point of view, i.e., in terms
of bifurcations, instabilities and topological changes in phase space.
In fact, the plain study of stationary properties of stochastic systems can 
already signal the occurrence of resonance phenomena and can contribute to 
uncover new mechanisms. Here, we are going to focus on the evaluation 
of the stationary probability distribution of the stochastic delay system, eq.(\ref{aa}). 
Before we \corr{are going to}{} investigate the impact of time delay let us first
briefly recall how this simple bifurcation structure is affected by noise, 
i.e., let us consider first the case $K=0$ and $D>0$.

There exist various nonequivalent notions of bifurcations in random systems.
The approach used predominately in the mathematical context, see, e.g., \cite{ARN03},
puts these notions on a rigorous basis. There are two types of stochastic bifurcations: phenomenological bifurcations (P-bifurcations) and dynamical bifurcations (D-bifurcations).
While the first type denotes a qualitative change of the shape of the distribution, the latter describes a change of stability for the trajectories. But such advanced theoretical concepts
are often difficult to investigate and almost impossible to apply in a plain 
experimental context. The complementary notion of noise-induced phase transitions
has less appeal from the rigorous point of view, but is straightforward to evaluate, 
see e.g. \cite{HOR84}. Here one solely focuses on the shape of the stationary distribution and
tracks, in particular, the number of maxima of such a distribution to detect a stochastic P-bifurcation. 
While such an approach depends on the coordinate system 
it has the nice feature that one can easily evaluate transitions even if
just an experimental time series is available. Here we follow the latter approach, which has been 
used in investigations of coherence resonance, see for instance \cite{ZAK13}. 

\begin{figure}[h!]

\includegraphics[width=0.4 \textwidth]{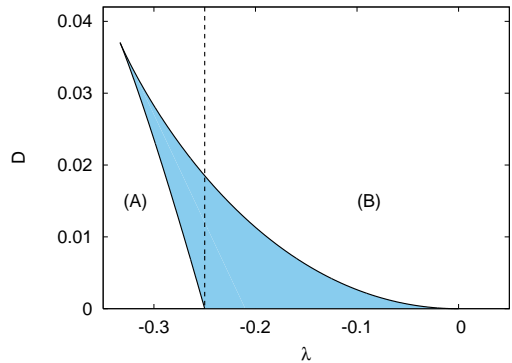}

\caption{Stochastic bifurcation diagram of the stochastic differential equation without delay eq.(\ref{aa}), corresponding to eq.(\ref{cac}) for $s=1$. 
The blue shaded area denotes the parameter values, where the probability distribution has a bimodal shape. 
The dashed line shows the border between the deterministic monostable (A) and the deterministic bistable (B) regime.} 
\label{figaa}
\end{figure}

If we consider the stochastic differential equation (\ref{aa})
for $K=0$, the stationary probability distribution can be written down if we 
solve the corresponding stationary Fokker-Planck equation \cite{RIS96}. 
In polar coordinates $z=r\exp(i\phi)$ the spherically symmetric probability distribution reads 
\begin{equation}\label{ca}
P(r)=N r \exp\left(\frac{r^2}{D}\left(\frac{\lambda}{2}+\frac{sr^2}{4}-\frac{r^4}{6}\right)\right),
\end{equation}
where $N$ is the normalisation constant.
Depending on the value of the two parameters, $\lambda$ and $D$, the distribution 
changes the shape from unimodal to bimodal behaviour, indicating a stochastic P-bifurcation
which is caused by a noise-induced limit cycle of finite amplitude. The transition
takes place if the exponent of the distribution $P$ develops an inflection point with regards to the
variable $r$, i.e., at $(\ln P)'=0$ and $(\ln P)''=0$
\begin{eqnarray}\label{caa}
-\frac{D}{r}-\lambda r-s r^3+r^5 &=& 0,\nonumber \\
\frac{D}{r^2}-\lambda-3 s r^2 + 5 r^4 &=& 0 \, .
\end{eqnarray}
These two conditions constitute a set of polynomial equations in the variable $r^2$, which reduces to an expression for the independent variable
\begin{equation}\label{cab}
r^2=-\frac{9D+\lambda s}{6 \lambda + 2 s^2} >0.
\end{equation}
Eq.(\ref{cab}) shows that only values for $\lambda < 0$ are allowed, otherwise one would face unphysical imaginary radii.
The resultant (in the mathematics literature called determinant) of the two polynomial equations \corr{}{(\ref{caa})} reads
\begin{equation}\label{cac}
\left(3D+\lambda s +\frac{2 s^3}{9}\right)^2=\left(\lambda s+\frac{2s^3}{9}\right)^2+\lambda^2\frac{(4 \lambda +s^2)}{3} \, .
\end{equation}
The bifurcation line in the $\lambda$-$D$ parameter plane which is explicitly
determined by eq.(\ref{cac})
has the typical triangular shape with a base covering the deterministic bistability range and
a cusp at $\lambda=-s^2/3$, $D=s^3/27$  (see figure \ref{figaa}). 

Hence, if we consider parameter
values in the subthreshold regime, $-s^2/3<\lambda<-s^2/4$, and gradually increase the strength
of the noise, $D$, noise-induced oscillations at finite amplitude appear at an optimal noise strength, which is
signalled by the second maximum in the stationary distribution.
If the noise exceeds a critical value, these oscillations are finally
wiped out again. Such features are quite common for noise-induced transitions 
in stochastic differential equations (see e.g. \cite{ZAK13}). 

\begin{figure}[h!]

\includegraphics[width=0.4 \textwidth]{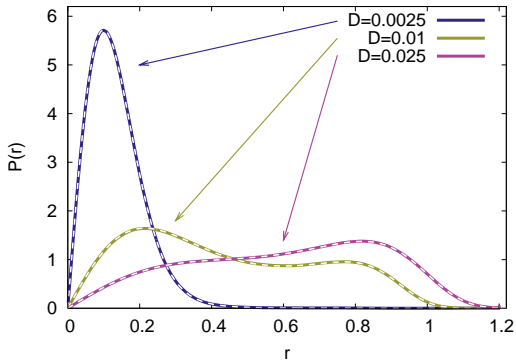}

\caption{Stationary amplitude probability distribution of the stochastic delay 
differential equation for $\omega=2\pi$, $\lambda=0.18$, $s=1$, $\tau=0.25$,
$K=0.5$ and three values of the
noise strength $D$ as obtained from numerical simulations of eq.(\ref{aa}) (solid lines).
Dashed lines (white) show the respective result of an analytical approximation scheme, eq.(\ref{ce}).
The analytical approximation is in excellent agreement with the
simulation results.}
\label{figa}
\end{figure}

It is the main purpose of the subsequent analysis to study how the classical
scenario is affected by time-delayed feedback. We first focus
on numerical simulations of eq.(\ref{aa}). We use a second order predictor-corrector scheme, along the lines of the Heun scheme (see \cite{KLO92} for 
related algorithms without time delay)  with stepsize of about $10^{-3}$.
Ensembles with about $10^9$ data points have been generated to
evaluate the stationary
behaviour. The data displayed in figure \ref{figa} show indeed that noise-induced
bifurcations between unimodal and bimodal distributions occur in the system
with delay when the noise strength is increased. Thus, the scenario known from the
stochastic differential equation persists in \corr{}{the} context of stochastic delay
differential equations.

\begin{figure}[h!]

\includegraphics[width=0.4 \textwidth]{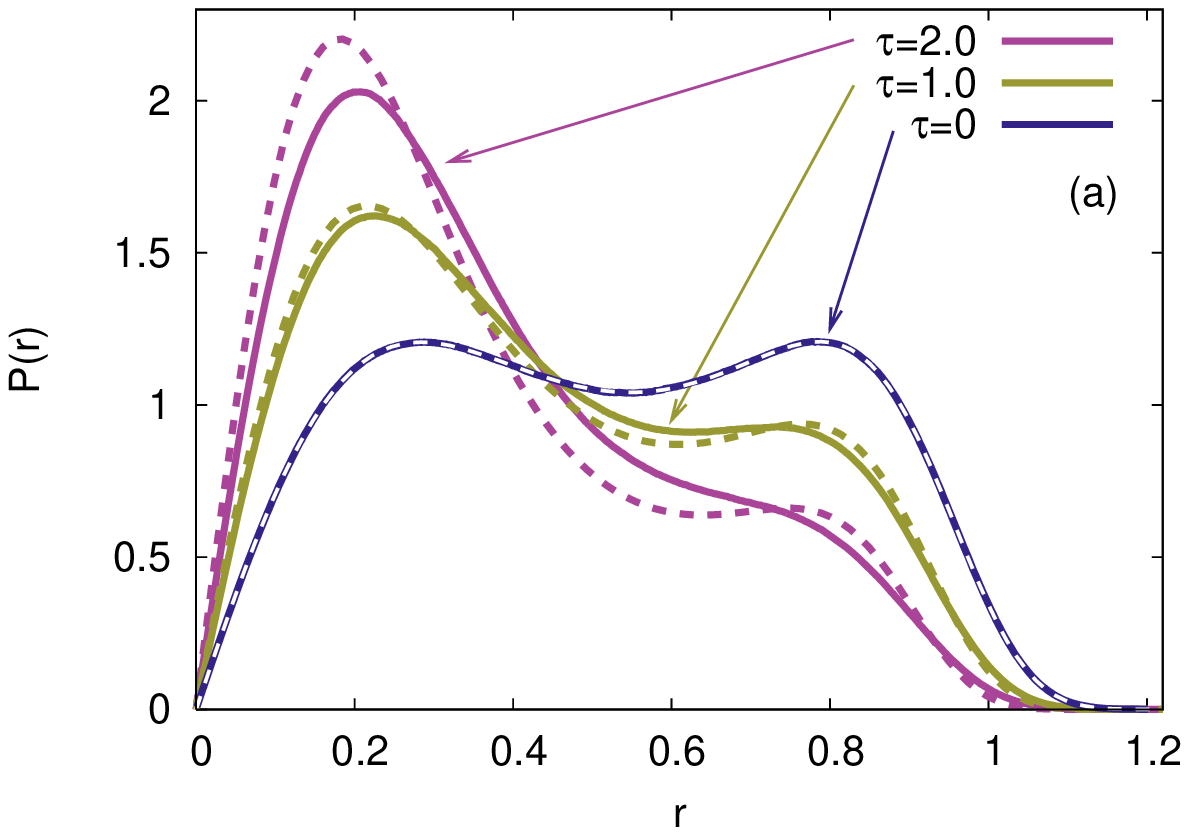}
\includegraphics[width=0.4 \textwidth]{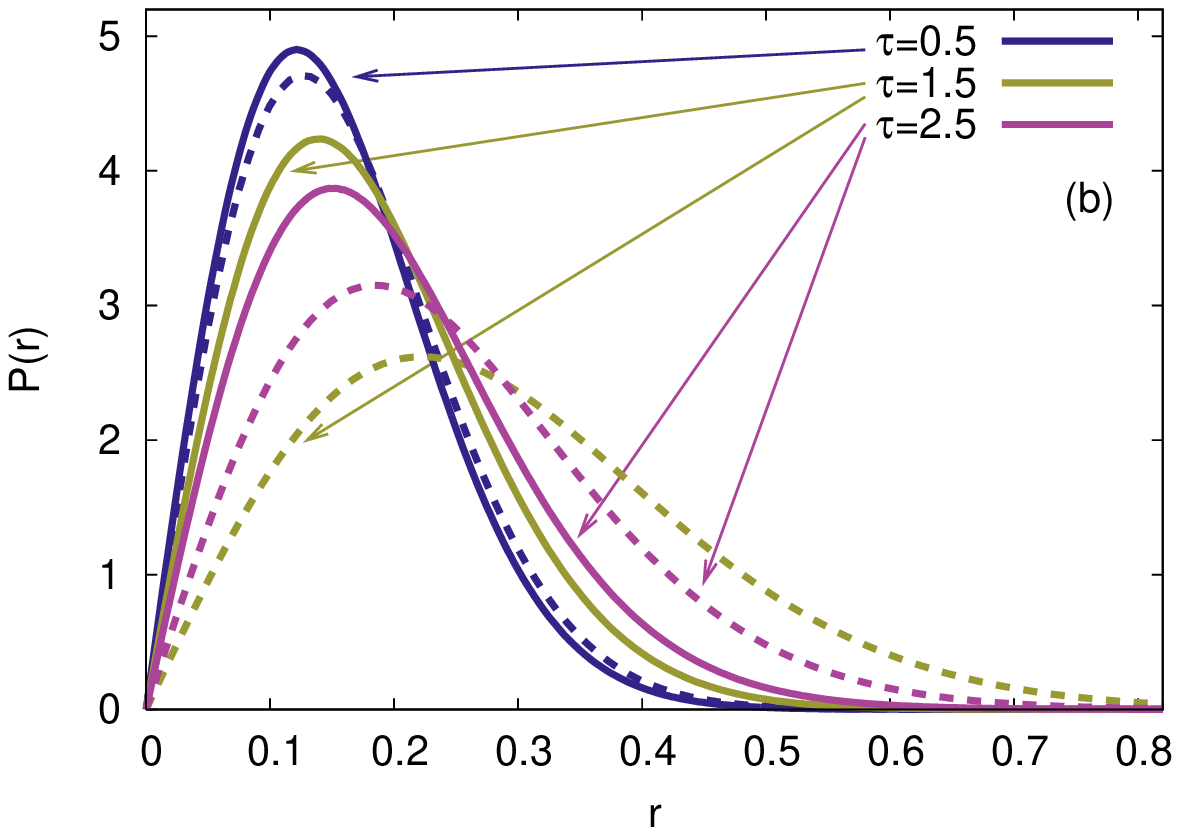}

\caption{Stationary amplitude probability distribution of the stochastic delay 
differential equation for $\omega=2\pi$, $\lambda=-0.26$, $s=1$, $D=0.015$,
$K=0.5$, and different values of time delay: (a) integer time delay, (b)
half integer time delay. Solid: results obtained from numerical simulations of
eq.(\ref{aa}), dashed: corresponding analytical approximation, eq.(\ref{ce}).
}
\label{figb}
\end{figure}

To illustrate the impact of time delay in a more systematic way
let us focus on a parameter setup which gives rise to a bimodal
distribution without delay and monitor how the shape of the distribution
changes with $\tau$. Figure \ref{figb} clearly demonstrates
that bimodality, i.e., the persistence of noise-induced oscillations
is supported by integer time delay (in units of the intrinsic frequency
$\omega$) (panel a) but that these oscillations are strongly suppressed at half
integer time delay (panel b). In that respect the feedback in eq.(\ref{aa})
shares some similarity with time-delayed feedback schemes (cf. e.g.
\cite{SCH07}). Thus there is a considerable impact on noise-induced bifurcations
when appropriate time-delayed feedback is applied.

To obtain further insight we clearly need a better understanding of the
dynamics of eq.(\ref{aa}) from an analytical point of view.
That goal is hampered by the non Markovian
character of the stochastic dynamics generated by eq.(\ref{aa}) (if considered
in the complex plane) and the infinite dimensionality of phase spaces
associated with functional differential equations \cite{HAL93}.
To make some progress we use the observation that the dynamical
phenomena we are interested in are at least to some extent related with
dynamical instabilities and bifurcations. Hence, adiabatic elimination schemes, the
reduction to effective degrees of freedom, or in formal terms a centre manifold reduction 
could be a way forward. For the mathematically inclined reader
some details of such formal centre manifold reductions, which also allow for the inclusion of higher order
perturbation series are given in appendix \ref{appb}. To illustrate the main idea
we present here a more intuitive but less rigorous account in terms of a 
multiple scale perturbation expansion (see, e.g., \cite{JUS98a} and as well \cite{JUS04} for such an approach 
in the context of time-delayed feedback control). 

We are going to base the analytical calculation on the
assumption that the parameters in eq.(\ref{aa}) are chosen such that
the deterministic part is close to the Hopf bifurcation, which we assume to occur for parameter
values $\lambda=\lambda_0$ and $K$. The characteristic equation, which follows from the linear
part of eq.(\ref{aa}) (see eq.(\ref{ba})), has a purely imaginary solution
$\Lambda=i \Omega_0$ and hence, reduces to (see as well eq.(\ref{bb}))
\begin{equation}\label{cb}
\lambda_0=K(1-\cos(\Omega_0 \tau)), \quad \Omega_0=\omega-K\sin(\Omega_0 \tau) \, .
\end{equation}
The linear deterministic part of eq.(\ref{aa}) induces a harmonic oscillation
with frequency $\Omega_0$, i.e., $z(t)=A\exp(i \Omega_0 t)$. 
We now consider parameter values close to such an instability.
For that purpose we introduce a formal small parameter $\varepsilon$
which measures the distance from the instability.
If scaled appropriately the time-dependent solution of the equation of motion
(\ref{aa}) is then given by an amplitude \corr{and timescale}{} 
modulated oscillation of the type
$z(t)=\varepsilon A(\varepsilon^4 t) \exp(i \Omega_0 t)$. The appropriate scaling 
which balances the impact of all the different contributions in eq.(\ref{aa})
at an order $\varepsilon^5$ reads $\lambda-\lambda_0=\delta \lambda
\rightarrow \varepsilon^4 \delta \lambda$, 
$s\rightarrow\varepsilon^2 s$, and $D\rightarrow \varepsilon^6 D$.
If we use a simple Taylor series expansion for the delayed feedback term, 
$A(\varepsilon^4(t-\tau))= A(\varepsilon^4 t)-\varepsilon^4 \tau A'(\varepsilon^4 t)+\ldots$ 
then eq.(\ref{aa}) at leading order $\varepsilon^5$ results in
\begin{eqnarray}\label{cc}
A'(\theta) &=& (\delta \lambda+s |A(\theta)|^2-|A(\theta)|^4)A(\theta) \\
& & - K \tau \exp(-i \Omega_0 \tau) A'(\theta) +\sqrt{2D}\exp(-i\Omega_0 t) \zeta(\theta), \nonumber
\end{eqnarray}
where $\theta=\varepsilon^4 t$ denotes the slow time scale (and $'$ the corresponding derivative).
If we recall that a phase transformation leaves an isotropic 
complex valued white noise invariant, we finally end up with 
\begin{eqnarray}\label{cd}
A'(\theta) &=& \frac{(\delta \lambda+s |A(\theta)|^2-|A(\theta)|^4)A(\theta)}{1+K \tau \exp(-i\Omega_0 \tau)}
 \nonumber \\
& & + \frac{\sqrt{2D} \zeta(\theta)}{|1+K \tau \exp(-i\Omega_0 \tau)|}.
\end{eqnarray}
Thus, close to an instability one is able to reduce the
stochastic delay dynamics, eq.(\ref{aa}), to the ordinary stochastic
differential equation (\ref{cd}) with Markovian property. 
Since eq.(\ref{cd}) has a phase symmetry,
it is rather straightforward to write down the 
spherically symmetric stationary probability distribution (see, e.g.,
\cite{RIS96})
\begin{eqnarray}\label{ce}
P(r)&=& N r \exp\left(\frac{r^2}{D_{eff}}\left(\frac{\delta \lambda}{2}+\frac{sr^2}{4}-\frac{r^4}{6}\right)\right), \nonumber \\
\delta \lambda &=& \lambda-K(1-\cos(\Omega_0 \tau)), \nonumber \\
D_{eff} &=& \frac{D}{1+K\tau \cos(\Omega_0 \tau)} \, .
\end{eqnarray}
Here we have again taken advantage of the spherical symmetry, 
which allows us to replace $A$ by $r\exp(i\phi)$.

Eq.(\ref{ce}) constitutes an analytical approximation for the stationary distribution
of the stochastic time delay system, eq.(\ref{aa}). The expression differs from
eq.(\ref{ca}), i.e. the stationary distribution of the system without time delay, by a
shift in the parameter $\lambda$ and a rescaling of the noise, see eq.(\ref{ce}).
The shift in the parameter $\lambda$ reflects the bifurcation structure of
the deterministic time delay system (see appendix \ref{appa}). In addition, and
somehow counterintuitively, the time delay has also an influence on the effective 
noise intensity in the system. Above all the impact of time delay can be made quantitative. 
The frequency $\Omega_0$ is determined by the implicit condition, eq.(\ref{cb}) as 
a function of $K$ and $\tau$.
A priori the expression eq.(\ref{ce}) is valid close to an instability. However, 
the numerical results clearly demonstrate that the analytic expression
is valid in a  surprisingly large parameter region. Figure \ref{figa} clearly shows that 
numerical simulations and the analytical approximation are virtually indistinguishable
for moderate values of the delay. Even for larger values of the delay, see figure \ref{figb},
where deviations become noticeable the analytic expression still captures quite well
the suppression of the bimodality of the stationary probability distribution at half integer delay times. 

Eq.(\ref{ce}) certainly fails if the effective noise strength changes
sign and becomes negative. The corresponding mechanism, i.e., the sign
of the expression $1+K\tau \cos(\Omega _0 \tau)$, can in fact be related to the
bifurcation diagram of the deterministic dynamical system,
where the sign of this expression determines the character of the underlying Hopf bifurcation
(see appendix \ref{appa} for the details). Interestingly, such a sub-supercritical transition 
was one of the key elements rectifying in misjudgements about time-delayed feedback control \cite{FIE07}.
Overall, some care has to be applied when one uses the analytical expression, 
eq.(\ref{ce}),
when $K \tau$  becomes larger than unity. In such cases it is easy to see that eq.(\ref{cb}) 
provides no longer a unique solution for $\Omega_0$.  A ``nontrivial'' value enters the 
expression eq.(\ref{ce}) for which the effective noise strength remains to be positive. These 
``nontrivial'' values for $\Omega_0$ are in fact directly related with the eigenvalue 
causing the Hopf instability on which the analytical approach has been based.
As can be seen from figure \ref{figb}, one again recovers quite a satisfactory analytical 
approximation even if the product $K \tau$ becomes larger than unity, and the 
approximation in fact seems to improve if one moves away from the ``critical'' case $K \tau=1$. 
In summary, the stationary probability distribution of the stochastic time delay system and its
analytical approximation reflects some of the fine structures of the underlying 
deterministic time delay dynamics.

\begin{figure}[h!]

\includegraphics[width=0.4 \textwidth]{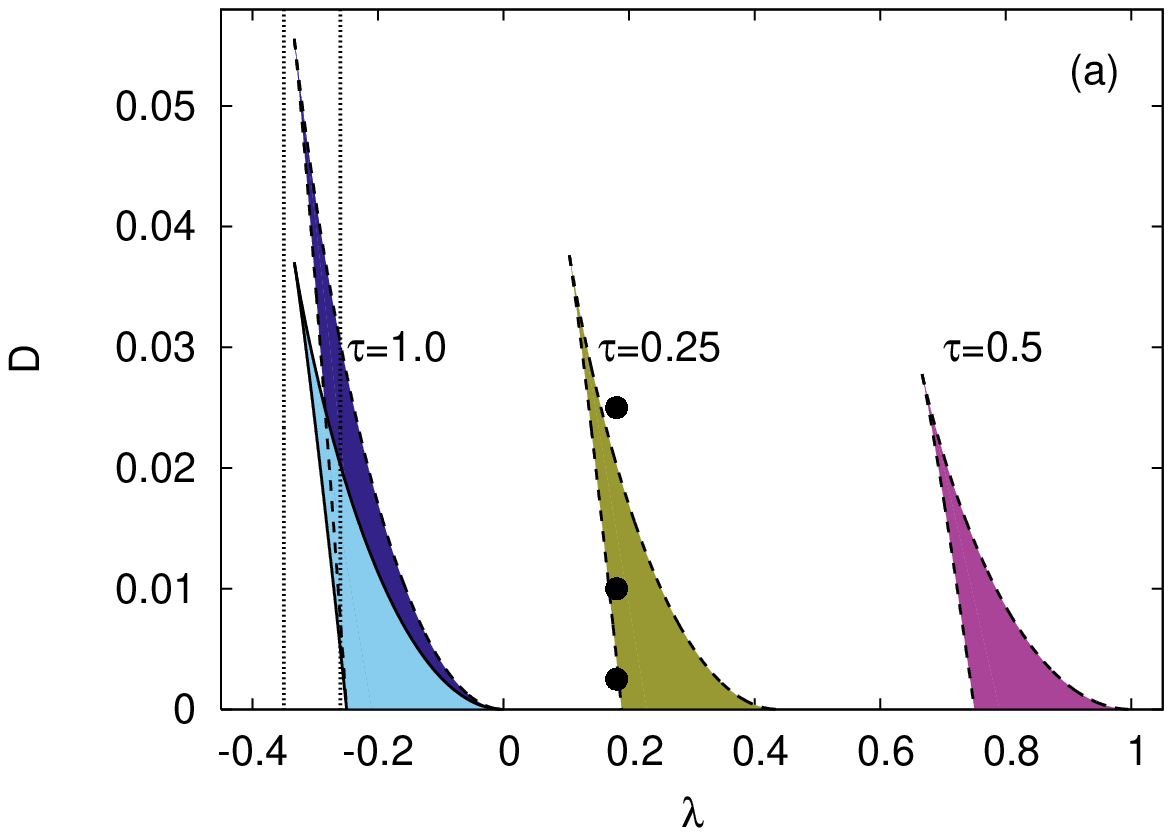}
\includegraphics[width=0.4 \textwidth]{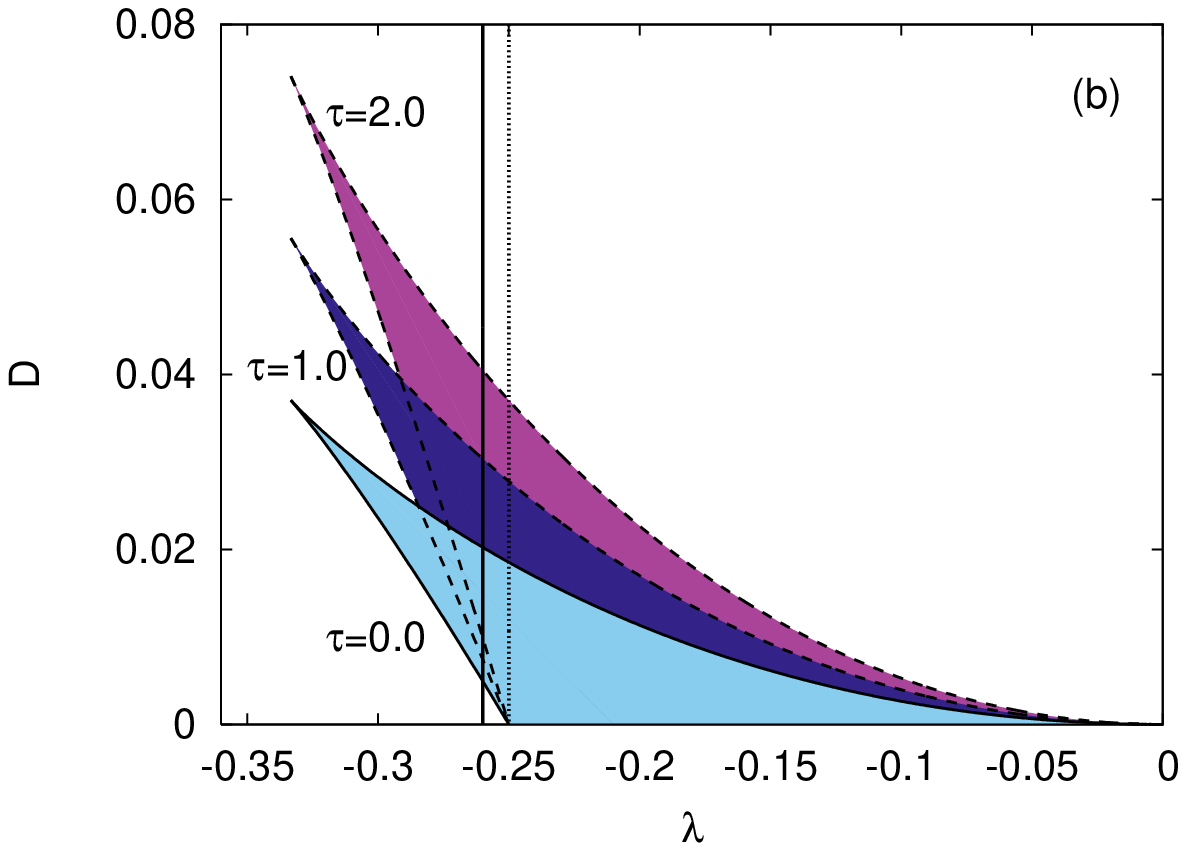}

\caption{Stochastic bifurcation diagram of the stochastic delay
differential equation (\ref{aa}) in the $(\lambda,D)$ parameter plane
for $s=1$, $K=0.5$, $\omega=2\pi$
\corr{}{as obtained from the stationary probability distribution eq.(\ref{ce}).
(a): 
Bifurcation diagram for small and moderate values of the delay.}
Solid/light: bimodal regime for $\tau=0$
(see figure \ref{figaa});
dashed/dark: time delay system with $\tau=0.25$ (bronze), $\tau=0.5$ (red) and $\tau=1.0$ (blue).
The filled circles are the parameter settings used
for figure \ref{figa}. The dotted vertical lines are the two paths used for the power
spectra shown in figure \ref{figf}.
\corr{}{(b): Detailed view for larger integer values of the delay.
Solid/light: deterministic system $\tau=0$;
dashed/dark: time delay system with $\tau=1.0$ (blue) and $\tau=2.0$ (red).
The dashed vertical line splits the figure up into the two deterministic regimes 
(compare figure \ref{figaa}). The vertical solid line shows the path in the
parameter plane, which is used in figure \ref{figh} to explore the
resonance characteristics of the correlation time.}
}
\label{figc}
\end{figure}

Since the analytic expression, eq.(\ref{ce}), captures quite well the
features of the stationary probability distribution, we may indeed base the computation
of the stochastic bifurcation diagram on the analytic approximation. As eq.(\ref{ce})
differs from eq.(\ref{ca}) just by a delay dependent rescaling of the parameters
the same is true for the stochastic bifurcation lines, see eq.(\ref{cac}).
Figure \ref{figc} shows the result of such a straightforward  analysis.
Because of the scale invariance of eq.(\ref{cac}) it is in fact sufficient to
constrain the analysis to the case $s=1$.
The most striking impact of the delay is a considerable shift of the
bifurcation lines which is mainly periodic with regards to the time delay, 
together with a mild scaling of the triangular region.

\section{Correlation properties and coherence resonance}\label{sec3}

Coherence resonance is a dynamical phenomenon, which manifests itself in a nonmonotonic dependence
of the correlation time upon the noise intensity, exhibiting a maximum at non-zero noise strength. 
There is also compelling evidence that
coherence resonance is indicated by multimodality of stationary probability distributions
and related stochastic bifurcations \cite{ZAK13}. The model, eq.(\ref{aa}), is a
suitable system to test such conjectures and to study in particular the impact
of time delay on such scenarios. For that purpose we evaluate the
autocorrelation function of the variable $x(t)=\mbox{Re}(z(t))$ and the corresponding power
spectral density $I(\Omega)$. To begin we focus on numerical simulations of
the stochastic time delay system eq.(\ref{aa}). We generate time traces of length
$2^9 \tau$ with step size $2^{-9} \tau$ resulting in a frequency resolution of about
$\Delta \Omega \simeq 10^{-2}$ with a cut-off frequency at about 
$\Omega_{max}\simeq 10^3$. The statistical average is taken
over an ensemble of $5000$ time traces. In order to discount for linear resonance phenomena
we monitor the spectrum normalised by the strength of the noise, $I(\Omega)/D$.
In this way only effects caused by nonlinearities will show up
in changes of the power spectrum.

\begin{figure}[h!]

\includegraphics[width=0.4 \textwidth]{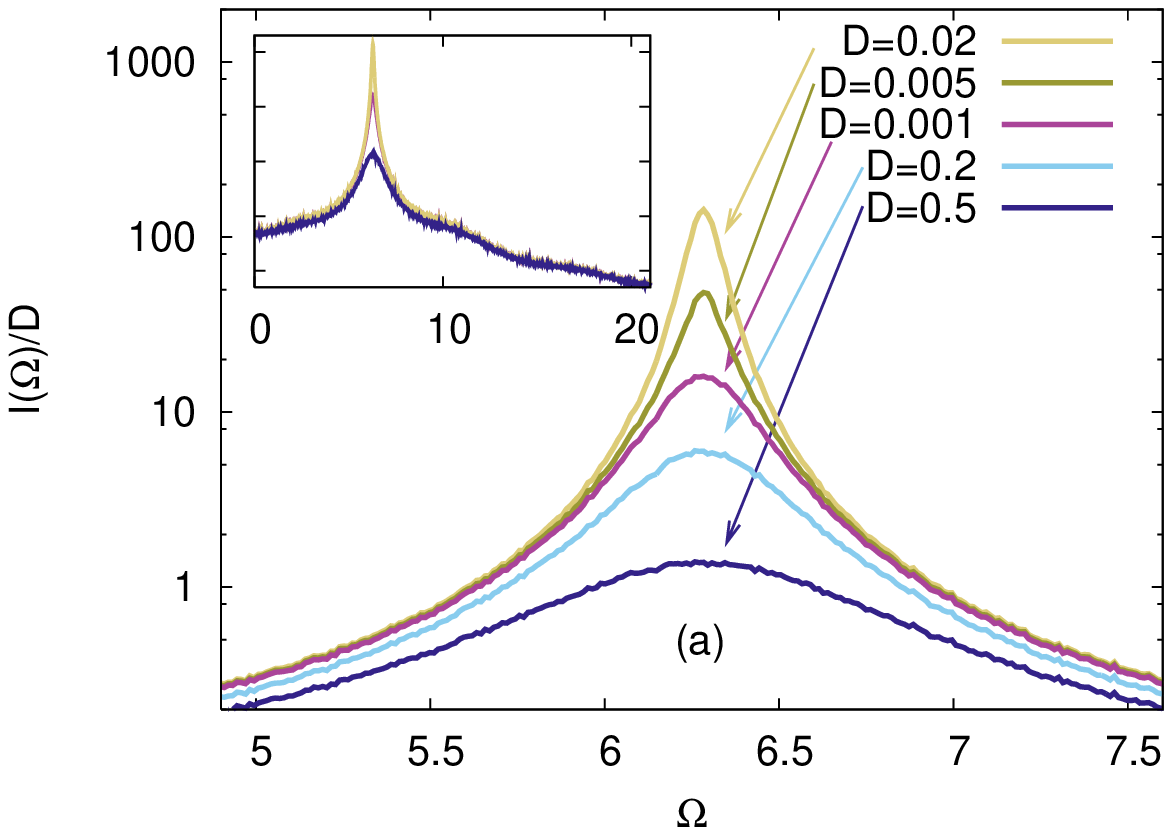}
\includegraphics[width=0.4 \textwidth]{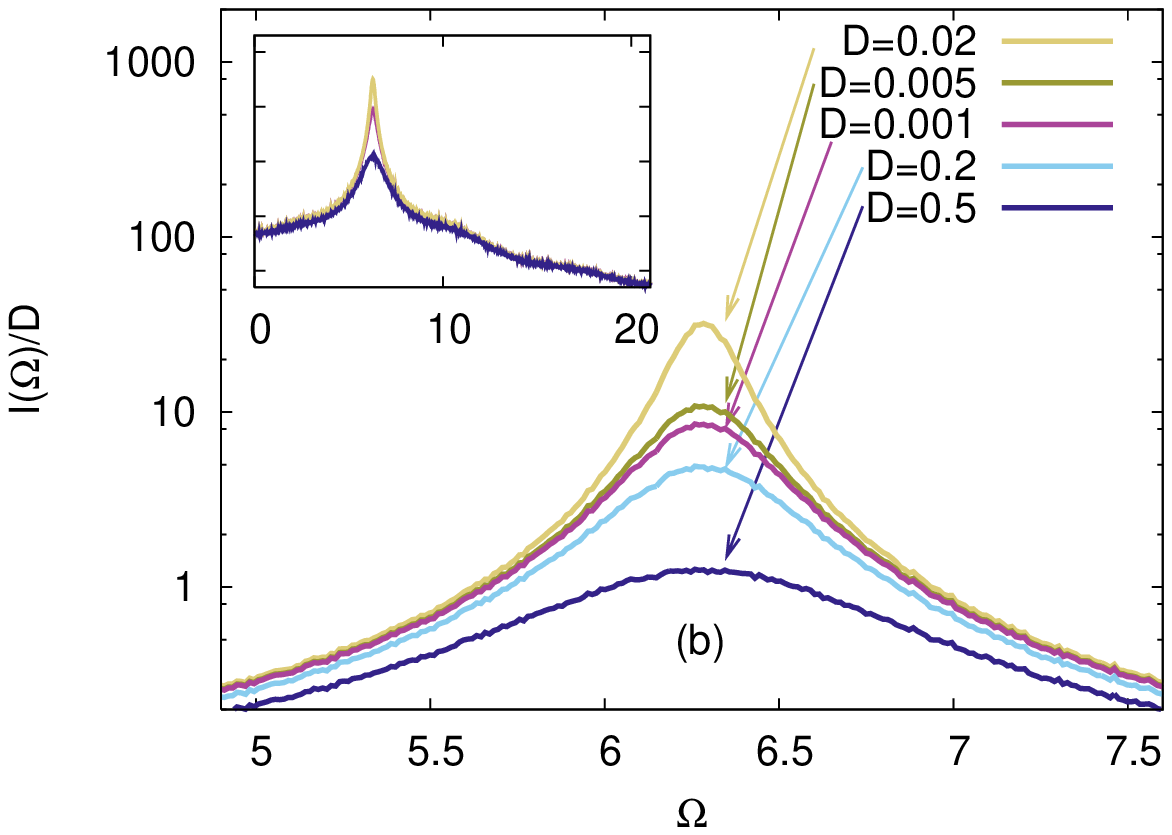}

\caption{Normalised power spectral density $I(\Omega)/D$ in the vicinity of the central peak
at $\Omega=2 \pi$, as obtained from a numerical simulation of eq.(\ref{aa}) for
increasing values of the noise intensity $D$, $\tau=1.0$, $\omega =2\pi$, $K=0.5$, $s=1$
and  two different values of $\lambda$ (a) $\lambda=-0.26$, (b) $\lambda=-0.35$
(see figure \ref{figc}). The insets show the structure of the power spectral densities
for $D=0.001$, $D=0.01$ and $D=0.1$ on  a larger scale.}
\label{figf}
\end{figure}

To demonstrate coherence resonance and its relation with stochastic bifurcations
we consider the dependence of the normalised spectrum on the noise strength for parameter 
variations, which do and which do not cross the stochastic bifurcation region, cf. figure
\ref{figc}. Figure \ref{figf}(a) shows pronounced coherence resonance for noise
amplitudes within the region of a bimodal stationary distribution. On increasing $D$
the additional power accumulates in the central peak, which increases by more than one order of
magnitude compared to the background spectrum at
an optimal noise amplitude $(D=0.02)$. On increasing $D$ further, the power then diffuses to the
background spectrum and the central peak diminishes. The other parts of the spectrum are
mainly unaffected by the change of the noise power. This coherence resonance phenomenon is not
strictly linked with stochastic bifurcations since the same feature, even though less
pronounced, is observed if we consider parameter variations which do not cross the
region of bimodal stationary probability distributions, see figure \ref{figf}(b).

\begin{figure}[h!]

\includegraphics[width=0.4 \textwidth]{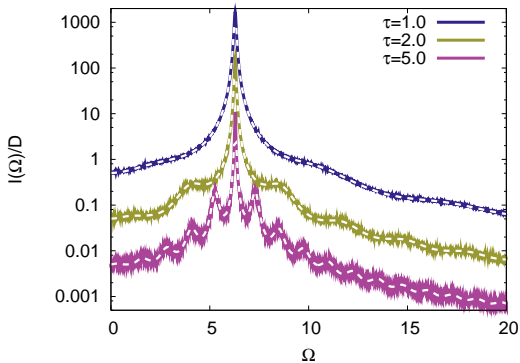}

\caption{Normalised power spectral density $I(\Omega)/D$ for
$D=0.015$, $\omega= 2\pi$, $K=0.5$, $\lambda=-0.26$, $s=1$ and 
different values of the delay.
Solid: numerical simulation of the nonlinear stochastic delay equation (\ref{aa}), 
dashed(white): corresponding analytical approximation according to eq.(\ref{en}).
For visibility the top/bottom spectrum is shifted by $\pm 10\mbox{dB}$,
respectively.}
\label{figg}
\end{figure}

As one can already guess from the structure of the bifurcation diagram of
the deterministic model (see appendix \ref{appa}, figure \ref{figd1}) 
one expects a considerable
impact of the time delay on the dynamics of the stochastic system. We are
mainly interested in changes of the coherence resonance phenomenon. In order
to focus on essential features we \corr{}{initially} consider integer delays where shifts
of the stochastic bifurcations are minimised, and on a noise strength
close to the optimal value. As visible from the data displayed in figure
\ref{figg}, an increase of the time delay leads to a moderate sharpening of the
central peak while on increasing the time delay further additional satellite
lines occur. Such a feature is quite common in time-delayed feedback 
systems, and can be qualitatively attributed to a frequency filtering caused by the
particular form of the time-delayed feedback term \cite{SCH04b}.

The numerical findings presented so far just give a glimpse of all the
phenomena that may appear in eq.(\ref{aa}). To gain deeper 
insight we need an analytic technique to approximate
correlations in nonlinear stochastic time delay systems. For that purpose
we adopt ideas that can be described as a variant of statistical linearisation
and which have proven to be quite successful for analytical approximations
of Fokker-Planck systems with nonlinear drift \cite{KOT86}. The main idea of the method,
as applied to the nonlinear stochastic time delay system, eq.(\ref{aa}), consists
in approximating the nonlinear contribution $s|z(t)|^2 z(t) - |z(t)|^4 z(t)$ by an effective 
linear term $\alpha z(t)$ resulting in a linear equation of motion
\begin{equation}\label{ea}
\dot{z}(t)=(\tilde{\lambda}+i\omega)z(t)-K(z(t)-z(t-\tau))+\sqrt{2D}\zeta(t),
\end{equation}
where 
\begin{equation}\label{eaa}
\tilde{\lambda} = \lambda + \alpha 
\end{equation}
denotes the rescaled bifurcation parameter. 
Contrary to a self-consistent mean field approach, which would
replace $|z(t)|^2$ and $|z(t)|^4$ by its average values assuming some underlying
Gaussian distribution (see e.g. \cite{POM05a}), statistical linearisation aims at an approximation 
which gives the best estimate of the nonlinear term by the linear one on average.
Hence, the value of the coefficient $\alpha$ is determined by minimising a suitable
norm
\begin{equation}\label{eb}
\left\langle \left|s|z(t)|^2 z(t) - |z(t)|^4 z(t) -\alpha z(t)\right|^2 \right\rangle \rightarrow \mbox{min.} \, .
\end{equation}
In this way we obtain
\begin{equation}\label{ec}
\alpha=\frac{s \left\langle |z(t)|^4 \right\rangle - \left\langle |z(t)|^6\right\rangle}{\left\langle |z(t)|^2 \right\rangle},
\end{equation}
where the averages are taken with regards to the stationary distribution.

The dynamics of the linear stochastic time delay system, eq.(\ref{ea}), which
provides an analytic approximation to the nonlinear dynamics of eq.(\ref{aa}), can be
solved in closed form (see, e.g. the seminal contribution \cite{KUE92}). While
the expression for the autocorrelation function
can be found in the literature we here provide the reader with all the main steps
of the calculation, to keep the presentation self-contained. We largely follow the reasoning of
\cite{AMA05}. Linear equations can be
conveniently solved in terms of eigenmodes. The characteristic equation corresponding to
eq.(\ref{ea}) reads
\begin{equation}\label{ed}
\Lambda=\tilde{\lambda}+i\omega-K(1-\exp(-\Lambda \tau)) \, .
\end{equation}
One can solve this transcendental equation by using the 
different branches of the Lambert W-function \cite{COR96}
\begin{eqnarray}\label{ee}
\Lambda_{\ell}=\frac{W_\ell[K\tau\exp(-(\tilde{\lambda}+i\omega-K)\tau)]}{\tau} +\tilde{\lambda}+i\omega -K, \nonumber \\
\ell \in \mathbb{Z}. \,
\end{eqnarray}
The time-dependent solution of eq.(\ref{ea}) is then expressed as a superposition
of eigenmodes
\begin{equation}\label{ef}
z(t)=\sum_\ell C_{\ell}(t),
\end{equation}
where the time evolution of the coefficients is governed by a variation of constants formula
\begin{equation}\label{eg}
\dot{C}_{\ell}(t) =\Lambda_{\ell} C_{\ell}(t)+\frac{\sqrt{2D}\zeta(t)}{N_{\ell}},
\end{equation}
with the weight 
\begin{equation}\label{efa}
N_{\ell}=1+K\tau \exp(-\Lambda_{\ell} \tau)
\end{equation}
following from the normalisation of eigenmodes. Solving eq.(\ref{eg})
the stationary solution follows from eq.(\ref{ef})
\begin{equation}\label{eh}
z(t)=\sqrt{2D} \int_0^\infty T(t') \zeta(t-t') dt',
\end{equation}
where we have introduced the propagator of the linear equation (\ref{ea})
\begin{equation}\label{ei}
T(t) = \sum_\ell \frac{\exp(\Lambda_{\ell} t)}{N_{\ell}}, \quad (t>0) \, .
\end{equation}
Using the correlation properties of the normalised complex valued noise,
$\langle \zeta(t) \zeta(t')\rangle=0$ and $\langle \zeta(t) \zeta^*(t')\rangle= 2 \delta(t-t')$, 
we easily conclude that the autocorrelation functions of
the real part $x(t)=(z(t)+z^*(t))/2$ and imaginary part $y(t)=(z(t)-z^*(t))/(2i)$ 
can be written as
\begin{equation}\label{ej}
\langle x(t) x(0)\rangle = \langle y(t) y(0)\rangle = \frac{1}{2}\mbox{Re} \langle z(t) z^*(0)\rangle \, .
\end{equation}
Using again the correlation properties of the noise the remaining autocorrelation
function follows from eq.(\ref{eh}) and eq.(\ref{ei}) after some short
algebra
\begin{equation}\label{ek}
\langle z(t) z^*(0)\rangle = 4 D \sum_{\ell \ell'} \frac{\exp(\Lambda_{\ell} t)}{N_{\ell} N_{\ell'}^*(-\Lambda_{\ell}-\Lambda_{\ell'}^*)} \, .
\end{equation}
For the power spectral density, i.e., for the Fourier transform 
of eq.(\ref{ej}) we finally arrive at 
\begin{equation}\label{el}
I(\Omega) = 2D \mbox{Re} \sum_{\ell \ell'} \frac{-2 \Lambda_{\ell}}{(\Lambda_{\ell}^2+\Omega^2)N_{\ell} N_{\ell'}^* (-\Lambda_{\ell} -\Lambda_{\ell'}^*)}.
\end{equation}
As expected, the power spectral density essentially consists of Lorentzian lines where the linewidth
is determined by the real part of the eigenvalues, eq.(\ref{ee}). One can express the series, eq.(\ref{el}), in closed analytic form.
For that purpose we recall that the propagator, eq.(\ref{ei}), obeys the deterministic equation 
of motion, eq.(\ref{ea}) with $D=0$, with initial condition $T(0)=1$ and $T(\theta)=0$, $-\tau<\theta<0$.
Hence, the Laplace transform of this solution reads
$(s-(\lambda+i\omega+\alpha-K)-K\exp(-s \tau))^{-1}$
\cite{BEL63},
which for the choice $s=-i\Omega$ results in the sum rule
\begin{eqnarray}\label{em}
\hat{T}(\Omega)&=&\sum_\ell \frac{1}{(i\Omega-\Lambda_{\ell}) N_{\ell}}  \\
&=& \frac{1}{i\Omega -(\tilde{\lambda}+i\omega -K)-K\exp(-i\Omega \tau)} \, . \nonumber
\end{eqnarray}
It is now rather straightforward to show that the sum in 
eq.(\ref{el}) can be essentially expressed as the absolute square of the Laplace transform,
eq.(\ref{em}), so that we arrive at the closed analytic formula
\cite{KUE92}
\begin{equation}\label{en}
I(\Omega)=D(|\hat{T}(\Omega)|^2 + |\hat{T}(-\Omega)|^2) \, .
\end{equation}

The data shown in figure \ref{figg} prove the accuracy of the analytical approximation, which is
able to reproduce the numerical results within 1\% of accuracy
over a rather wide generic range of parameter values. 
Within our level of resolution the analytic and 
the numerical values can hardly be distinguished and one can safely use the analytic formula to
estimate coherence resonance signatures in our model. To some extent the success of the
approximation depends crucially on the correct estimate of the effective
coefficient by eq.(\ref{ec}).

\begin{figure}[h!]

\includegraphics[width=0.4 \textwidth]{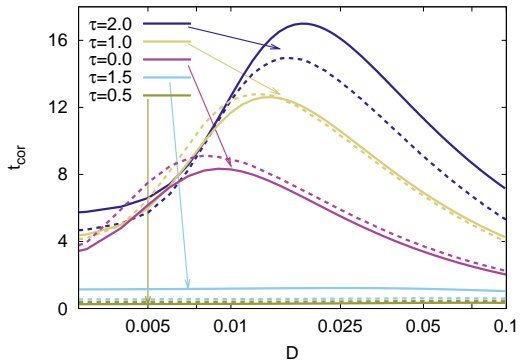}

\caption{Influence of the time delay on the correlation time $t_{cor}$. The dashed lines correspond to numerical simulations
of eq.(\ref{aa}), the solid lines show the approximation, calculated from eq.(\ref{ep}) with $\ell=\ell'=0$. 
Parameters: $\omega = 2\pi$, $\lambda = -0.26$, $K = 0.5$, $s=1$.
}
\label{figh}
\end{figure}

To \corr{show}{demonstrate} the modulation of coherence resonance caused by the time delay, we use the correlation time, 
which in a linear approximation is proportional to the inverse of the width $\Gamma$ of the main spectral peak. 
\corr{The correlation time is introduced in \cite{STR63} as}{Alternatively, the correlation time may be introduced  as in  \cite{STR63} by integrating over the
absolute value of a normalised autocorrelation function, i.e.}
\begin{equation}\label{eo}
 t_{cor} = \frac{1}{\langle x(0) x(0)\rangle}\int_0^{\infty}{|\langle x(t)x(0)\rangle| dt} .
\end{equation}
\corr{is the normalised correlation function of the real part of the state 
$z(t)$. 
The function $\Psi _{xx}(t)$ can be derived by using eq.(\ref{ej}) and eq.(\ref{ek}) and the normalisation is evaluated by the correlation function at $t=0$. 
The triangle inequality and the filling factor 
$\frac{1}{\pi}\int_{-\pi/2}^{\pi/2}\cos (\phi)d\phi = \frac{2}{\pi}$ is used 
for estimating the integral in eq.(\ref{eo}).  
We obtain a closed-form expression for the correlation time, which reads}{The
correlation function can be evaluated using  eqs.(\ref{ej}) and eq.(\ref{ek}).
To deal with the modulus we approximate the fast oscillating part of the autocorrelation function by
$\pi^{-1}\int_{-\pi/2}^{\pi/2}\cos (\phi)d\phi = 2/\pi$ \cite{SCH04b}. Thus we arrive at}
\begin{equation}\label{ep}
t_{cor} = \frac{2}{\pi \langle x(0) x(0) \rangle}
\sum_{\ell \ell^{'}}\frac{1}{\left|N_\ell N_{\ell^{'}}^*(-\Lambda _\ell - \Lambda _{\ell^{'}}^*)\right| 
\text{Re}(-\Lambda_\ell) } . 
\end{equation}
For the numerical \corr{simulations, we calculated time series of length $10^4$, step size $10^{-3}$, and took every $20$th data point. 
For averaging, $300$ time series were simulated, each for $100$ different values of the noise intensity}{evaluation of eq.(\ref{eo}) 
we simulated an ensemble of 300 time traces of length $10^4$ with about $5\times 10^5$ data points, to
compute the autocorrelation function and the integral of its absolute value. The simulations were repeated for 100 different values of the noise intensity in the
interval $D \in [0.001, 0.1]$. The comparison between the simulation and the
analytical estimate is shown in figure \ref{figh}.
Note that for the analytical results 
we only take the main branch of the Lambert W-function 
($\ell=\ell'=0$) in eq.(\ref{ep}) into account.} 
For small time delays the main branch dominates the dynamics and 
adding more branches does not cause
significant changes of the result. Our approximation is in good agreement 
with the numerics for \corr{the}{}integer delay times.

The impact of the delay on the correlation time is 
\corr{shown}{clearly visible} in figure \ref{figh}. 
\corr{If correlation time achieves a maximum}{The maximum value of the correlation time} at a certain optimal noise intensity \corr{}{signals} 
\corr{the system features}{} coherence resonance. 
As one would expect from the results in section II, coherence resonance is suppressed for half-integer delay times and enhanced for integer ones. 
Furthermore, an increase of the integer delays shifts the optimal noise intensity to higher values and the correlation becomes stronger. 
The shift of the optimal noise intensity can be explained by the fact that the regime of bimodal probability distribution is changed by the delay time, which is shown in figure \ref{figc}. 
For a fixed value of $\lambda$ a higher value of the noise is necessary to reach the regime of bimodality, see figure \ref{figc}(b). 

The enhancement and the suppression of the correlation time is caused by the stability of the deterministic focus: it shows
non-monotonic behaviour as a function of $\tau$, \corr{}{cf. figure \ref{figd1} (App.A) 
for the bifurcation diagram of the underlying deterministic system.}
\corr{It}{The focus} becomes more stable for half integer delays and less stable for integer delays.
It is easier to excite the system with the noise in the case of a less stable focus and therefore the correlation time has a higher
value with delayed feedback in the system. This was already shown in \cite{JAN03}: for a less stable focus, the coherence of oscillations
is higher. \corr{}{However, this interpretation based on the linear properties of the trivial fixed point does not constitute the complete picture. 
The increase in the signal-to-noise ratio and the related increase in correlation time is entirely due to the nonlinear features of the dynamical system, i.e., due to the emerging saddle-node bifurcation.}

\begin{figure}[h!]

\includegraphics[width=0.4 \textwidth]{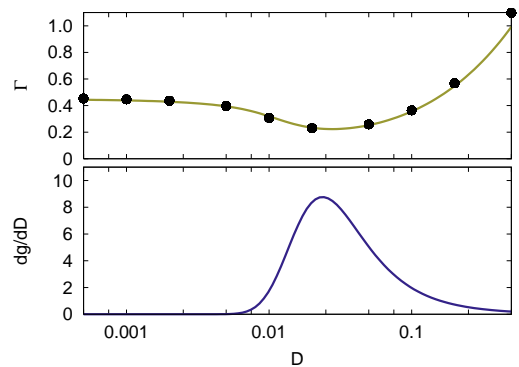}

\caption{Signature of coherence resonance of eq.(\ref{aa}) outside the bimodal regime at
$\lambda=-0.35$ (left dotted vertical line in Fig.\ref{figc}a), $\omega=2 \pi$, $K=0.5$, $\tau=1.0$, and $s=1$
(see figure \ref{figc}).
Upper panel: Linewidth $\Gamma$ of the central peak of the power spectrum 
as a function of the noise intensity $D$ as obtained from the analytic approximation, 
eq.(\ref{ee}) with $\alpha$ being computed from the analytic expression (\ref{ce}) 
(solid line). Symbols mark the values as obtained from the numerical simulations 
(see figure \ref{figf}(b)). Lower panel: dependence of the ghost weight derivative $dg/dD$ on the
noise intensity $D$.}
\label{figi}
\end{figure}

Noise-induced bifurcations and the development of multimodality have been considered as a key characteristic of coherence resonance, but such a viewpoint is not entirely satisfactory. 
In \cite{ZAK13} it is pointed out that the highest degree of regularity at a certain optimal noise intensity, i.e., pronounced coherence resonance, is provided by a stationary distribution with bimodal shape. 
But outside of the bimodal regime less pronounced coherence resonance can still be observed. Hence, it is tempting to figure out features within the stationary distribution that can be used to \corr{explain the mechanism of}{detect} 
coherence resonance. 
Within our model \corr{}{the mechanism of} coherence resonance is related with the ghost of the stable periodic orbit that is created at $\lambda=-s^2/4$ and $|z|=\sqrt{s/2}$ in the deterministic system without time delay. 
To measure the significance of such a phase space structure 
we suggest to consider the cumulative probability of states having amplitudes beyond the critical radius $|z|=\sqrt{s/2}$ at which the pair of stable and unstable periodic orbits is created
\begin{equation}\label{fa}
g(D)=\int_{\sqrt{s/2}}^\infty P(r) dr \, .
\end{equation}
Consequently, the quantity $g(D)$ is the weight of the probability distribution above the radius of the ghost limit cycle induced by noise. Therefore, we call it ghost weight. 
The ghost weight can even be evaluated by analytical means using eq.(\ref{ce}).
It is essentially a monotonically increasing function as the noise intensity is spreading out the probability in phase space. 
The resonance characteristics, i.e., the dynamical impact of the phase space structure close to the saddle-node bifurcation point, is indicated by the increment of the function, i.e., by the sensitivity of $g(D)$ with respect to $D$. 
Indeed, the derivative $dg/dD$ is able to show quite accurately the resonance, see figure \ref{figi}. 

\begin{figure}[h!]

\includegraphics[width=0.4 \textwidth]{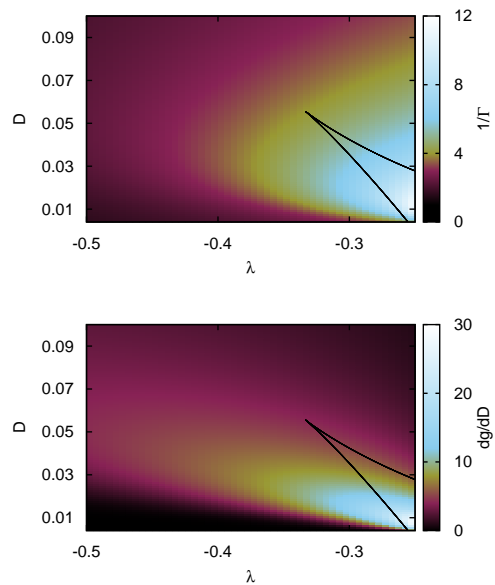}

\caption{Dependence of the correlation time, i.e., of the inverse linewidth $\Gamma$
(top panel) and of the ghost weight derivative (lower panel) on $\lambda$ and $D$
as obtained from the analytic approximations, eqs.(\ref{ce}) and
(\ref{ee}) for the system, eq.(\ref{aa}) with  $\omega=2 \pi$, $K=0.5$, $\tau=1.0$,
and $s=1$. The line indicates the result of the noise-induced bifurcation
analysis, i.e., the region of a bimodal stationary distribution, see figure \ref{figc}.
}
\label{figj}
\end{figure}

That property seems to be quite generic, as it holds in a rather large part of the parameter region, see figure \ref{figj}. 
While the numerical values of the correlation time $t_{cor}$ and of $dg/dD$ certainly 
differ (they do not even have the same units) the resonance characteristics, i.e., the maximum 
visible in the ridge of the two functions coincide quite nicely. 
The derivative of the ghost weight together with the correlation time can be used to detect and hence, provides an explanation of the mechanism of coherence resonance outside the regime of bimodality. 

In summary, the techniques behind eq.(\ref{en})
provide powerful tools to estimate correlation
properties of nonlinear stochastic time delay systems.

\section{Conclusion}\label{sec4}

We have presented an extensive study of coherence resonance 
in non-excitable systems with a particular
focus on the impact of time-delayed feedback. For that purpose we 
introduced a paradigmatic model system, i.e., a Hopf normal form
subjected to noise and a time-delayed feedback coupling. That 
\corr{set up}{setup} allowed us 
to study in detail the coherence resonance phenomenon related to the saddle-node bifurcation
of periodic orbits which is connected with a subcritical Hopf bifurcation. 
Given that normal forms are able to capture features of general dynamical systems
we propose that our findings have a rather large range of applicability.

We have applied analytical tools to investigate
stochastic time delay equations and to confirm the numerical findings of our
model. For the stationary probability distribution a centre manifold analysis 
has been developed to predict and confirm the shape of the
stationary probability distribution with a particular focus on detecting stochastic
bifurcations. While a priori such analytic approximations are only
valid close to bifurcation points we find that our expression, eq.(\ref{ce}), 
is able to capture even quantitatively the behaviour of the stationary state 
in a large region of the parameter space. Time delay leads to a characteristic modulation and scaling
of the stochastic bifurcation pattern as visible in figure \ref{figc}.
That pattern can be captured by parameters renormalised by the time delay, 
see eq.(\ref{ce}).

To investigate correlation properties we have used the fact that
linear stochastic time delay systems, like their deterministic counterpart, 
admit a complete analytic solution. Adapting the concept of stochastic linearisation, 
which is known to deliver rather accurate analytic approximations for
Fokker-Planck equations with nonlinear drift, we have derived 
closed expressions for the autocorrelation function as well as the power spectral density.
Such an approximation scheme captures the numerical findings quite well, in fact,
within 1\% of the numerical values. There is excellent agreement between the numerical
results obtained from the power spectral density and the analytical estimates (see, e.g., figure \ref{figg}). 
Thus we are able to evaluate correlation
properties in detail. 

The maximum of a
correlation time is often used as an indicator for \corr{}{}coherence resonance. 
The correlation time may be defined in two different ways: either a direct calculation from the
autocorrelation function, or the inverse of the 
linewidth in the power spectral density. The latter is
analytically accessible from the eigenvalues, eq.(\ref{ee}). 
The correlation time shows a maximum, i.e., the linewidth shows a minimum,
which indicates coherence resonance even if the parameter setting
does not correspond to a bimodal stationary probability distribution, see figure \ref{figi}. Therefore the ghost weight, which is calculated from the probability distribution, but does not depend on its shape, has been introduced \corr{for the explanation}{as a signature} of coherence resonance. In particular, it allows to draw an analogy between the mechanism of coherence resonance in excitable systems based on two competing time-scales and in non-excitable systems, where this phenomenon occurs due to the competition between the noisy focus and the ghost of the limit cycle.
Hence, coherence resonance and noise-induced bifurcation are two slightly different facets
of stochastic time delay systems.

The investigation of our simple model systems and the tools and signatures
that have been developed in this context may help to broaden our understanding of
coherence resonance in stochastic delay systems, and to open up insight into the dynamical interplay between stochastic inputs and time
delayed interactions.

\begin{acknowledgements}
This work was supported by DFG in the framework of SFB 910.
\end{acknowledgements}

\appendix

\section{Bifurcation analysis of the deterministic model}\label{appa}

\begin{figure}[h!]

\includegraphics[width=0.4 \textwidth]{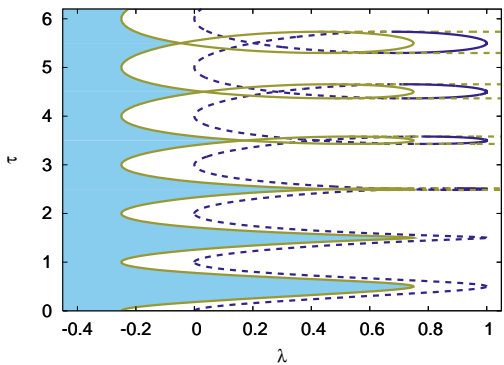}

\caption{Hopf curves $\tau _h$ and saddle-node bifurcation curves $\tau _{sn}$ in 
the $(\lambda,\tau)$ plane,  corresponding to eq.(\ref{bc}) and eq.(\ref{bg}), 
for $\omega = 2 \pi$, $K=0.5$, $s=1$, and  $n=0,..,6$.
Blue (black): Hopf bifurcation curves $\tau _h$ (solid: supercritical bifurcation, dashed: subcritical bifurcation), 
bronze (grey): saddle-node bifurcation of limit cycles curves $\tau _{sn}$ (solid: rigid bifurcation, dashed: non-rigid bifurcation). 
The light blue shaded area indicates the region with stable trivial fixed point and without any harmonic periodic orbit.
It is of interest for coherence resonance (see figure \ref{figk}).}
\label{figd1}
\end{figure}

The bifurcation analysis of the deterministic model, i.e., eq.(\ref{aa}) with $D=0$, can 
be performed to a large extent by analytical means. The linear stability of the trivial fixed point 
$z=0$ is  obviously governed by the characteristic equation \cite{HOE05}
\begin{equation}\label{ba}
\Lambda=\lambda+i\omega -K(1-\exp(-\Lambda \tau)) \, .
\end{equation}
If the eigenvalue becomes purely imaginary, $\Lambda=i\Omega$, a Hopf bifurcation takes place. Thus,
eq.(\ref{ba}) yields (see also eq.(\ref{cb}))
\begin{eqnarray}\label{bb}
\Omega &=& \omega - K\sin (\Omega \tau), \label{bba}\\
\lambda &=& K(1-\cos(\Omega \tau)) \label{bbb} \quad .
\end{eqnarray}
We discuss the
bifurcation diagram in a $(\lambda,K)$ parameter plane, 
for fixed value of the delay $\tau$. We first 
uncover the structure of the various bifurcation lines in the
$(\lambda,K)$ parameter plane and translate the results to the
$(\lambda,\tau)$ plane at the end of this appendix (see figure \ref{figd1}).

\begin{figure}[h!]

\includegraphics[width=0.4 \textwidth]{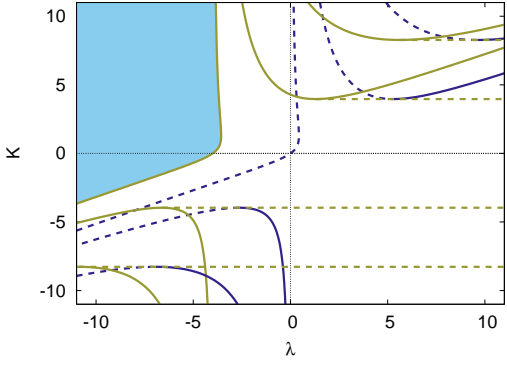}

\caption{Bifurcation diagram of the deterministic part of eq.(\ref{aa}) for
$D=0$, $\tau=0.75$ and $s=4$, due to eq.(\ref{bd}) and eq.(\ref{bh}). Blue: Hopf bifurcation lines;
solid/dashed: super/subcritical bifurcation. Bronze: saddle-node bifurcation of
periodic solutions; solid/dashed: rigid/non-rigid bifurcation (see figure \ref{fige}). 
The blue shaded area indicates the region with stable trivial fixed point and without any periodic orbit.
It is of interest for coherence resonance (see figure \ref{figk}).}
\label{figd}
\end{figure}

Eq.(\ref{ba}) yields for the bifurcation lines the parametric representation
\begin{eqnarray}\label{bd}
K &=& \frac{\omega - \Omega}{\sin (\Omega \tau)},  \\
\lambda &=& K(1-\cos(\Omega \tau)),\quad
\Omega\tau\in (n\pi,(n+1)\pi) \, . \nonumber 
\end{eqnarray}
Generically, eq.(\ref{bd}) results in a single monotonic branch and a nested structure
of parabolic branches (see figure \ref{figd}). As for the direction of the bifurcation
one may compute the change of eigenvalue with respect to $\lambda$. Implicit
differentiation of eq.(\ref{ba}) results in
\begin{equation}\label{be}
\mbox{Re}\left.\left(\frac{d\Lambda}{d\lambda}\right)\right|_{\Lambda=i\Omega}=\frac{1+K \tau \cos(\Omega \tau)}{|1+K \tau \exp(i \Omega \tau)|^2} \, .
\end{equation}
Since the extrema of the bifurcation lines, eq.(\ref{bd}),  are determined by
the condition $1+K\tau \cos(\Omega \tau)=0$
we conclude that the real parts become positive when crossing 
the left part of one of the parabolic branches and negative when crossing 
the right branch (see figure \ref{figd}). Finally, to evaluate the type of Hopf bifurcation one
would need to evaluate the cubic coefficient of the corresponding normal form.
It is rather straightforward but slightly tedious to show that the real part
of the cubic coefficient is given by $s(1+K \tau \cos(\Omega \tau))$ (see
appendix \ref{appb}). Hence, the Hopf bifurcations on the single monotonic branch as
well as on the left part of a parabolic branch are subcritical while
those on the right part are supercritical (see figure \ref{figd}). At the extrema
a higher order codimension transition appears from sub- to supercritical behaviour.

\begin{figure}[h!]

\includegraphics[width=0.4 \textwidth]{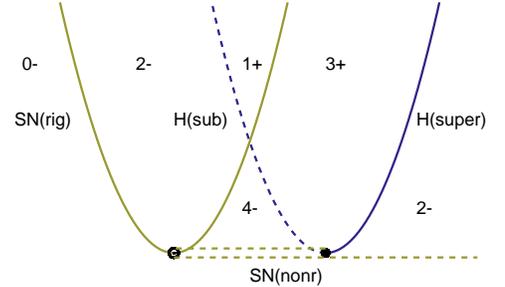}

\caption{Scheme of the Hopf and the saddle-node bifurcation lines
in the bifurcation diagram of eq.(\ref{aa}), see figure \ref{figd}.
Blue (black) (dashed/solid) sub/supercritical Hopf bifurcation of the trivial fixed point.
Bronze (grey): rigid (solid) and non-rigid (dashed) saddle-node bifurcation lines. The numbers indicate the 
harmonic solutions created in the saddle-node and Hopf bifurcations,
the signs show the change of relative stability of the trivial fixed point
due to the Hopf bifurcation.}
\label{fige}
\end{figure}

Such higher order codimension points are associated with saddle-node bifurcations of
periodic states. To compute these bifurcations it is sufficient to focus on harmonic
solutions of eq.(\ref{ba}), $z(t)=A\exp(i\alpha t)$, say with $A>0$, and to determine 
the parameter regions where such solutions exist. Eq.(\ref{ba}) yields
\begin{eqnarray}\label{bf}
-|A|^4+s |A|^2 &=& -\lambda +K(1-\cos(\alpha \tau)),\label{bfa}\\
\alpha &=& \omega -K\sin(\alpha \tau) \, . \label{bfb}
\end{eqnarray}
In fact, eq.(\ref{bfa}) can be written as
$(|A|^2-s/2)^2 =\lambda +s^2/4 -K(1-\cos(\alpha \tau)) \geq 0$ so that a pair of stable and unstable periodic orbits 
is generated when the right hand side of the inequality vanishes.
Thus, eqs.(\ref{bfa}) and (\ref{bfb}) result
in the parametric representation of a saddle-node bifurcation line
\begin{eqnarray}\label{bh}
K=\frac{\omega -\alpha}{\sin(\alpha \tau)}, \quad \lambda=-s^2/4+K(1-\cos(\alpha \tau)),  \\
\alpha\tau\in (n\pi,(n+1)\pi) \,  \nonumber .
\end{eqnarray}
It is obvious from the previous discussion that the pair of periodic orbits is created on
increasing $\lambda$. By comparing eqs.(\ref{bd}) and (\ref{bh}) we conclude
that any branch of the Hopf bifurcation line induces a corresponding saddle-node bifurcation 
line which is obtained by shifting the Hopf bifurcation line by
$-s^2/4$ (see figures \ref{figd} and \ref{fige}). These saddle-node bifurcation lines do not terminate, and thus
do not link up with the sub-supercritical codimension-two transition mentioned previously. The pair of orbits which is 
created in the saddle-node bifurcation have the common period $\alpha$, mainly due to the high symmetry of our
equations of motion, see eq.(\ref{bfb}). Thus, to label this type of bifurcation we call it
a rigid saddle-node bifurcation. 

The second mechanism for the creation of a pair of harmonic solutions, i.e., 
for a saddle-node bifurcation, is caused by eq.(\ref{bfb}). On changing the
parameter $K$ a real pair of $\alpha$ values can be created resulting in two
harmonic solutions of different period. The corresponding condition 
for this type of bifurcation follows from the derivative of eq.(\ref{bfb}) and
obviously reads
\begin{equation}\label{bi}
1+K\tau \cos(\alpha \tau)=0 \, .
\end{equation}
\corr{Eq.(\ref{bf}) determines}{Eqs.(\ref{bf}) and (\ref{bi}) determine} horizontal straight bifurcation lines which precisely hit the
the extrema of the Hopf and the rigid saddle-node bifurcation lines (see figures \ref{figd}, \ref{fige}). Since we still have to ensure that eq.(\ref{bfa}) gives real solutions for the amplitude
the line does not extend to infinity. In fact, the bifurcation line determined by eq.(\ref{bi})
starts at the sub-supercritical transition with a small value
for $|A|$. It extends to the left until $\lambda=-s^2/4$ where it hits
the extremum of the corresponding rigid saddle-node bifurcation line, and where eq.(\ref{bfa}) provides 
a doubly degenerate solution. Hence, on increasing $\lambda$ we obtain another
branch of the saddle-node bifurcation line with large amplitudes $|A|\ge s/2$
which then extends from  $\lambda=-s^2/4$ to infinity (see  figure \ref{fige}). 
As this saddle-node bifurcation results in a pair of periodic orbits which attain
different periods we label the saddle-node bifurcation as non-rigid.

After having clarified the structure of the bifurcation diagram in the $(\lambda,K)$ parameter plane
(see figure \ref{figd}) it is straightforward to translate the
results to a $(\lambda,\tau)$ plane which is relevant to visualise the impact
of time delay. In particular, using trigonometric identities
for the elimination of $\Omega$ from eqs.(\ref{bba}) and (\ref{bbb})
we even obtain an explicit expression for the Hopf bifurcation line \cite{HOE05}
\begin{equation}\label{bc}
 \tau _h = \frac{\pm \arccos \left((K-\lambda)/K\right) + 2\pi n}{\omega \mp \sqrt{2K\lambda - \lambda ^2}},~n\in \mathbb{N}.
\end{equation}
In a similar way (cf. eqs.(\ref{bfa}) and (\ref{bfb}))
we derive a closed analytic formula for the rigid saddle-node bifurcation line
\begin{equation}\label{bg}
 \tau _{sn} = \frac{\pm \arccos \left((K-\lambda-s^2/4)/K\right) + 2\pi n}{\omega \mp 
\sqrt{(8K - 4\lambda - s^2)(4\lambda + s^2 )}/4},~n\in \mathbb{N}.
\end{equation}
Figure \ref{figd1} shows the result of this analysis for the parameter setting
used in sections \ref{sec2} and \ref{sec3}, and thus allows 
a clear understanding of the impact of time delay.

\section{Centre manifold reduction of the stochastic delay differential equation}\label{appb}

To perform a centre manifold reduction we consider a parameter value $\lambda=\lambda_0+\delta \lambda$ 
close to a Hopf bifurcation, i.e., $\lambda_0$ obeys eq.(\ref{ba}) with eigenvalue $\Lambda=i \Omega_0$
(cf. eq.(\ref{cb}) as well). The formal calculation follows standard procedures \cite{HAL93}.
We here follow the approach proposed in \cite{AMA05}. Using the condition on the parameter $\lambda$ we rewrite eq.(\ref{aa}) as
\begin{equation}\label{da}
\dot{z}(t)=(\lambda_0-K) z(t)+K z(t-\tau) + f(t),
\end{equation}
where the inhomogeneous part, given by
\begin{equation}\label{db}
f(t)=(\delta \lambda + s |z(t)|^2-|z(t)|^4) z(t)+\sqrt{2D}\zeta(t)
\end{equation}
contains all the contributions of the unfolding which are supposed to be all of the 
same small order.
To perform a centre manifold reduction recall that the phase space of
a delay differential equation is given by a function space, here on $[-\tau,0]$.
Such a feature is captured by introducing the notation $Z_t(\theta)=z(t+\theta)$ with
the delay variable $\theta \in [-\tau,0]$. Since the solution of the linear part of eq.(\ref{da})
reads $Z_t(\theta)=\exp(i \Omega_0 t) \exp(i \Omega_0 \theta)$ the centre manifold
is tangential to $V(\theta)=\exp(i\Omega_0 \theta)$ and can be written as
\begin{equation}\label{dc}
z(t+\theta)= C(t) \exp(i \Omega \theta) + R(t,\theta)\, .
\end{equation}
Here, $C(t)$ denotes the complex coordinate on the centre manifold and $R(t)$ captures the higher
order terms of an expansion in $C(t)$. To define the coordinate 
on the centre manifold uniquely  one needs to impose a constraint on the higher order terms.
It is quite convenient to work with an orthogonality condition with regard to
the adjoint eigenmode of the linear system \cite{AMA05}. Such a constraint results in the
condition
\begin{equation}\label{dd}
0=R(t,0)+ K \int_{-\tau}^0 \exp(-i\Omega_0(\theta+\tau)) R(t,\theta) d \theta \, .
\end{equation}
Combining eqs.(\ref{dc}) and (\ref{dd}) we obtain for the coordinate
\begin{equation}\label{de}
N C(t)= z(t)+K\int_{t-\tau}^t \exp(i \Omega_0(t-\theta-\tau)) z(\theta) d\theta
\end{equation}
with the normalisation $N=1+K \tau \exp(-i\Omega_0 \tau)$. Taking the time derivative and
using eq.(\ref{da}) we arrive at the equation of motion on the centre manifold
\begin{equation}\label{df}
\dot{C}(t)=i\Omega_0 C(t)+f(t)/N \, .
\end{equation}
If we evaluate the expression (\ref{db}) on the centre manifold (\ref{dc}) to lowest
order we thus end up with (cf. eq.(\ref{cc}))
\begin{eqnarray}\label{dg}
\dot{C}(t) &=& i\Omega_0 C(t) +\frac{(\delta \lambda+s|C(t)|^2-|C(t)|^4) C(t)}{N}\, \nonumber \\
& & +\frac{\sqrt{2D} \zeta(t)}{N}.
\end{eqnarray}

\bibliographystyle{prsty-fullauthor}

\bibliography{ref}

\end{document}